\documentclass[
final            
  ]
  {aipproc}
\usepackage{amsfonts}

\layoutstyle{8x11double}

\newcommand{\ZZ}{\mathbb{Z}}
\newcommand{\RR}{\mathbb{R}}
\newcommand{\CC}{\mathbb{C}}
\newcommand{\HH}{\mathbb{H}}
\newcommand{\TT}{\mathbb{T}}
\newcommand{\calA}{\mathcal{A}}

\newcommand*{\Approx}[1]{\stackrel{#1}{\approx}}

\newcommand{\DeclareMathOperator}[2]{\def#1{\mathop{\mathrm{#2}}\nolimits}}

\let\Re\relax\DeclareMathOperator{\Re}{Re}
\let\Im\relax\DeclareMathOperator{\Im}{Im}

\newcommand{\pt}{\mathrm{pt}}
\DeclareMathOperator{\mod}{mod}

\DeclareMathOperator{\sgn}{sgn}

\DeclareMathOperator{\Pf}{Pf}
\DeclareMathOperator{\GL}{GL}

\DeclareMathOperator{\Cliff}{Ciff}

\newcommand{\spinup}{{\uparrow}}
\newcommand{\spindown}{{\downarrow}}

\newcommand*{\Bar}[1]{\bar{#1}}
\newcommand*{\V}[1]{\mathbf{#1}}
\newcommand*{\Vr}[1]{\mathbf{r}_{#1}}

\begin{document}

\title{Periodic table for topological insulators and superconductors}


\author{Alexei Kitaev}{
address={California Institute of Technology, Pasadena, CA 91125, U.S.A.}}

\begin{abstract}
Gapped phases of noninteracting fermions, with and without charge conservation
and time-reversal symmetry, are classified using Bott periodicity. The
symmetry and spatial dimension determines a general universality class, which
corresponds to one of the $2$ types of complex and $8$ types of real Clifford
algebras.  The phases within a given class are further characterized by a
topological invariant, an element of some Abelian group that can be $0$,
$\mathbb{Z}$, or $\mathbb{Z}_2$. The interface between two infinite phases
with different topological numbers must carry some gapless mode. Topological
properties of finite systems are described in terms of $K$-homology. This
classification is robust with respect to disorder, provided electron states
near the Fermi energy are absent or localized. In some cases (e.g., integer
quantum Hall systems) the $K$-theoretic classification is stable to
interactions, but a counterexample is also given.
\end{abstract}

\maketitle

The theoretical study~\cite{KaneMele0,KaneMele1,HgTe0} and experimental
observation~\cite{HgTe1} of the quantum spin Hall effect in 2D systems,
followed by the discovery of a similar phenomenon is 3
dimensions~\cite{MooreBalents,Roy,FuKaneMele,FuKane-BiSb,BiSb}, have generated
considerable interest in topological states of free electrons. Both kinds of
systems are time-reversal invariant insulators. More specifically, they
consist of (almost) \emph{noninteracting} fermions with a \emph{gapped energy
spectrum} and have both the time-reversal symmetry ($T$) and a $U(1)$ symmetry
($Q$). The latter is related to the particle number, which is conserved in
insulators but not in superconductors or superfluids.  Topological phases with
only one of those symmetries, or none, are also known.  Such phases generally
carry some gapless modes at the boundary.\footnote{In contrast, strongly
correlated topological phases (with anyons in the bulk) may not have gapless
boundary modes\cite{BravyiKitaev}.}

The classification of gapped free-fermion systems depends on the symmetry and
spatial dimension. For example, two-dimensional insulators without $T$
symmetry are characterized by an integer $\nu$, the quantized Hall
conductivity in units of $e^2/h$. For systems with discrete translational
symmetry, it can be expressed in terms of the band structure (more exactly,
the electron eigenstates as a function of momentum); such an expression is
known as the TKNN invariant~\cite{TKNN}, or the first Chern number. A similar
topological invariant (the $k$-th Chern number) can be defined for any even
dimension $d$. For $d=0$, it is simply the number of single-particle states
with negative energy ($E<E_F=0$), which are filled with electrons.

However, the other three symmetry types (no symmetry, $T$ only, or both $T$
and $Q$) do not exhibit such a simple pattern. Let us consider systems with no
symmetry at all.  For $d=0$, there is a $\ZZ_2$ invariant: the number of
electrons $(\mod 2)$ in the ground state.  For $d=1$, a system in this
symmetry class, dubbed ``Majorana chain'', also has a $\ZZ_2$ invariant, which
indicates the presence of unpaired Majorana modes at the ends of the
chain~\cite{Majorana}.  But for $d=2$ (e.g., a $p_x+ip_y$ superconductor), the
topological number is an integer though an even-odd effect is also
important~\cite{ReadGreen,hexagon}.

$T$-invariant insulators have an integer invariant (the number of
particle-occupied Kramers doublet states) for $d=0$, no invariant for $d=1$,
and a $\ZZ_{2}$ invariant for $d=2$~\cite{KaneMele0,KaneMele1} and for
$d=3$~\cite{MooreBalents,Roy,FuKaneMele}. 3D crystals (i.e., systems with
discrete translational symmetry) have an additional $3\ZZ_{2}$ invariant,
which distinguishes so-called ``weak topological insulators''.

With the exception just mentioned, the topological numbers are insensitive to
disorder and can even be defined without the spectral gap assumption, provided
the eigenstates are localized. This result has been established rigorously for
integer quantum Hall systems~\cite{Bellissard,ASB,BES-B}, where the invariant
$\nu$ is related to the index theory and can be expressed as a trace of a
certain infinite operator, which represents the insertion of a magnetic flux
quantum at an arbitrary point. Its trace can be calculated with sufficient
precision by examining an $l$-neighborhood of that point, where $l$ is the
localization length. A similar local expression for the $\ZZ_2$ invariant of a
1D system with no symmetry has been derived in Appendix~C of
Ref.~\cite{hexagon}; it involves an infinite Pfaffian or determinant.

\begin{table}
\vtop{\hbox{\begin{tabular}[t]{ccccc}
\hline
$q$ & $\pi_{0}(C_{q})$& $d=1$ & $d=2$ & $d=3$ \\
\hline
$0$ &  $\ZZ$ & & (IQHE) &\vspace{1pt}\\
$1$ &  $0$ & & &\\
\hline
\end{tabular}}\vskip7pt
\hbox{\parbox{5.3cm}{\textbf{Above:} insulators without time-reversal symmetry
(i.e., systems with $Q$ symmetry only) are classified using complex
$K$-theory.\\[7pt] \textbf{Right:} superconductors/superfluids (systems with
no symmetry or $T$-symmetry only) and time-reversal invariant insulators
(systems with both $T$ and $Q$) are classified using real $K$-theory.}}}
\hskip20pt
\begin{tabular}[t]{ccccc}
\hline
$q$ & $\pi_{0}(R_q)$ & $d=1$ & $d=2$ & $d=3$\vspace{1pt}\\
\hline
$0$ &  $\ZZ$ & &
\tabcolsep=0pt\begin{tabular}{c} no symmetry\\
($p_x+ip_y$, e.g., SrRu) \end{tabular}& 
\tabcolsep=0pt\begin{tabular}{c} $T$ only\\ 
($^{3}$He-$B$)\end{tabular}\vspace{7pt}\\
$1$ &  $\ZZ_2$ &
\tabcolsep=0pt\begin{tabular}{c} no symmetry\\
(Majorana chain)\end{tabular} &
\tabcolsep=0pt\begin{tabular}{c} $T$ only\\
$\bigl((p_x\!+\!ip_y)\spinup+(p_x\!-\!ip_y)\spindown\bigr)$\end{tabular} &
\tabcolsep=0pt\begin{tabular}{c}$T$ and $Q$\\ (BiSb)
\end{tabular}\vspace{7pt}\\
$2$ &  $\ZZ_2$ &
\tabcolsep=0pt\begin{tabular}{c} $T$ only\\
((TMTSF)${}_{2}$X)\end{tabular} &  
\tabcolsep=0pt\begin{tabular}{c} $T$ and $Q$\\ (HgTe)
\end{tabular} &\vspace{7pt}\\
$3$ &  $0$ &  $T$ and $Q$ &  &\vspace{1pt}\\
$4$ &  $\ZZ$ &  &  &\vspace{1pt}\\
$5$ &  $0$ &  &  &\vspace{1pt}\\
$6$ &  $0$ &  &  &\vspace{1pt}\\
$7$ &  $0$ &  &  &\kern-7pt no symmetry\kern-3pt\\
\hline &&&&\\
\end{tabular}
\caption{Classification of free-fermion phases with all possible combinations
of the particle number conservation ($Q$) and time-reversal symmetry
($T$). The $\pi_{0}(C_q)$ and $\pi_{0}(R_{q})$ columns indicate the range of
topological invariant. Examples of \emph{topologically nontrivial} phases are
shown in parentheses.}
\label{tab_periodic}
\end{table}

In this paper, we do not look for analytic formulas for topological numbers,
but rather enumerate all possible phases. Two Hamiltonians belong to the same
phase if they can be continuously transformed one to the other while
maintaining the energy gap or localization; we will elaborate on that
later. The identity of a phase can be determined by some local probe. In
particular, the Hamiltonian around a given point may be represented (in some
non-canonical way) by a mass term that anticommutes with a certain Dirac
operator; the problem is thus reduced to the classification of such mass
terms.

Prior to this work, there have been several results toward unified
classification of free-fermion phases. Altland and Zirnbauer~\cite{AZ}
identified 10 symmetry classes of matrices,\footnote{These classes are often
associated with random matrix ensembles, but the symmetry pertains to concrete
matrices rather than the probability measure.} which can be used to build a
free-fermion Hamiltonian as a second-order form in the annihilation and
creation operators, $\hat{a}_j$ and $\hat{a}_j^{\dag}$. The combinations of
$T$ and $Q$ make 4 out of 10 possibilities. However, the symmetry alone is
only sufficient to classify systems in dimension $0$. For $d=1$, one may
consider a zero mode at the boundary and check whether the degeneracy is
stable to perturbations. For example, an unpaired Majorana mode is stable. In
higher dimensions, one may describe the boundary mode by a Dirac operator and
likewise study its stability. This kind of analysis has been performed on a
case-by-case basis and brought to completion in a recent paper by Schnyder,
Ryu, Furusaki, and Ludwig~\cite{SRFL}. Thus, all phases up to $d=3$ have been
characterized, but the collection of results appears irregular.

A certain periodic pattern for $\ZZ_2$ topological insulators has been
discovered by Qi, Hughes, and Zhang~\cite{classInsulators}. They use a
Chern-Simons action in an extended space, which includes the space-time
coordinates and some parameters. This approach suggests some operational
interpretation of topological invariants and may even work for interacting
systems, though this possibility has not been explored. In addition, the
authors mention Clifford algebras, which play a key role in the present paper.

\begin{table}
\vtop{\hbox{\begin{tabular}[t]{ccccc}
\hline
$q \bmod 2$ & Classifying space $C_q$ & $\pi_{0}(C_q)$\\
\hline
$0$ & $\bigl(U(k+m)/(U(k)\times U(m))\bigr)\times\ZZ$ & $\ZZ$\vspace{1pt}\\
$1$ & $U(n)$ & $0$\\
\hline
\end{tabular}}\vskip4pt
\hbox{\parbox{7.8cm}{\textbf{Above:} The classifying space $C_{0}$
parametrizes Hermitian matrices $X$ with $\pm1$ eigenvalues.\, $C_{q}$ is the
$q$-th loop space of $C_{0}$; it parametrizes such matrices $X$ that
anticommute with $q$ Clifford generators.\\[5pt] \textbf{Right:} Similar
classification for real symmetric matrices.}}}
\hspace{20pt}
\begin{tabular}[t]{cccc}
\hline
$q\bmod 8$ & Classifying space $R_q$ & $\pi_{0}(R_q)$\\
\hline
$0$ & $\bigl(O(k+m)/(O(k)\times O(m))\bigr)\times\ZZ$ & $\ZZ$\vspace{1pt}\\
$1$ & $O(n)$ & $\ZZ_2$\vspace{1pt}\\
$2$ & $O(2n)/U(n)$ & $\ZZ_2$\vspace{1pt}\\
$3$ & $U(2n)/Sp(n)$ & $0$\vspace{1pt}\\
$4$ & $\bigl(Sp(k+m)/(Sp(k)\times Sp(m))\bigr)\times\ZZ$ & $\ZZ$\vspace{1pt}\\
$5$ & $Sp(n)$ & $0$\vspace{1pt}\\
$6$ & $Sp(n)/U(n)$ & $0$\vspace{1pt}\\
$7$ & $U(n)/O(n)$ & $0$\\
\hline
\end{tabular}
\caption{Bott periodicity in complex and real $K$-theory. (The parameters
$k,m,n$ should be taken to infinity.)}
\label{tab_Bott}
\end{table}

We report a general classification scheme for gapped free-fermion phases in
all dimensions, see Table~\ref{tab_periodic}. It actually consists of two
tables. The small one means to represent the aforementioned alternation in
TR-broken insulators (a unique trivial phase for odd $d$ vs. an integer
invariant for even $d$). The large table shows a period~8 pattern for the
other three combinations of $T$ and $Q$. Note that phases with the same
symmetry line up diagonally, i.e., an increase in $d$ corresponds to a step up
$(\mod 8)$. ($T$-invariant 1D superconductors were studied in
Ref.~\cite{midgap}. The $(p_x\!+\!ip_y)\spinup+(p_x\!-\!ip_y)\spindown$ phase
was proposed in Refs.~\cite{classSuper,Roy1,SRFL}; the last paper also
describes an integer invariant for $^{3}$He-$B$.) The $2+8$ rows (indexed by
$q$) may be identified with the Altland-Zirnbauer classes arranged in a
certain order; they correspond to 2 types of complex Clifford algebras and 8
types of real Clifford algebras. Each type has an associated \emph{classifying
space} $C_q$ or $R_q$, see Table~\ref{tab_Bott}. Connected components of that
space (i.e., elements of $\pi_{0}(R_q)$ or $\pi_{0}(C_q)$) correspond to
different phases. But higher homotopy groups also have physical meaning. For
example, the theory predicts that 1D defects in a 3D TR-broken insulator are
classified by $\pi_1(C_{1})=\ZZ$.

The $(\mod 2)$ and $(\mod 8)$ patterns mentioned above are known as \emph{Bott
periodicity}; they are part of the mathematical subject called
\emph{$K$-theory}. It has been applied in string theory but not so much in
condensed matter physics. One exception is Ho\v{r}ava's work~\cite{Horava} on
the classification of stable gapless spectra, i.e., Fermi surfaces, lines, and
points. In this paper, we mostly use results from chapters~II--III of
Karoubi's book~\cite{Karoubi}, in particular, the relation between the
homotopy-theoretic and Clifford algebra versions of $K$-groups (a variant of
the Atiyah-Bott-Shapiro construction~\cite{AtiyahBottShapiro}).

\section{Some examples}

To get a glimpse of the mathematical structure underlying the topological
classification, we consider a second-order transition between two phases,
where the energy gap vanishes at some value of parameters. In this case, the
low-energy Fermi modes typically have a Dirac spectrum, and the phases differ
by the sign of the mass term.

Let us begin with the simplest example, the Majorana chain~\cite{Majorana}.
This model has one spinless Fermi mode per site, but the number of particles
is not conserved, which calls for the use of \emph{Majorana operators}:
\begin{equation}
\hat{c}_{2j-1}=\hat{a}_j+\hat{a}_j^\dag,\quad
\hat{c}_{2j}=\frac{\hat{a}_j-\hat{a}_j^\dag}{i}\quad
(j=1,\dots,n).
\end{equation}
By convention, operators acting in the the Fock space (as opposed to the mode
space) are marked with a hat. The Majorana operators are Hermitian and satisfy
the commutation relations $\hat{c}_l\hat{c}_m+\hat{c}_m\hat{c}_l=2\delta_{lm}$;
thus, $\hat{c}_{1},\dots,\hat{c}_{2n}$ may be treated on equal footing. (But
it is still good to remember that $\hat{c}_{2j-1}$ and $\hat{c}_{2j}$ belong
to the same site $j$.) The advantage of the Majorana representation is that
all model parameters are real numbers.

A general free-fermion Hamiltonian for non-conserved particles has this form:
\begin{equation}\label{quadH}
\hat{H}_{A} = \frac{i}{4}\,\sum_{j,k} A_{jk}\hat{c}_{j}\hat{c}_{k},
\end{equation}
where $A$ is a real skew-symmetric matrix of size $2n$. The concrete model is
this:
\begin{equation}
\hat{H}=\frac{i}{2}
\left(u\sum_{l=1}^{n}\hat{c}_{2l-1}\hat{c}_{2l}
+v\sum_{l=1}^{n-1}\hat{c}_{2l}\hat{c}_{2l+1}\right).
\end{equation}
At the transition between `the ``trivial phase'' ($|u|>|v|$) and the
``topological phase'' ($|u|<|v|$), there are two counterpropagating gapless
modes. They may be represented by two continuous sets of Majorana operators,
$\hat{\eta}_j(x)$ ($j=1,2$). The effective Hamiltonian near the transition
point has this form:
\begin{equation}
\hat{H}=\frac{i}{2}\int\hat{\eta}^{T}\!
\left(\begin{array}{@{}cc@{}}\partial & m\\
-m & -\partial\end{array}\right)
\hat{\eta}\:dx,\qquad
\hat{\eta}=\left(\begin{array}{@{}c@{}}
\hat{\eta}_1\\
\hat{\eta}_2
\end{array}\right),
\end{equation}
where $m\sim u-v$. Thus, we need to study the Dirac operator
$D=\gamma\partial+M$, where $\gamma=\sigma^z$ and $M=m\,i\sigma^y$. If $m$
gradually varies in space and changes sign, e.g., $m(x)=-ax$, the Dirac
operator has a localized null state, which corresponds to an unpaired Majorana
mode in the second quantization picture. The existence of the true null state
is a subtle property, but it has a simple semiclassical analogue: a continuous
transition between a positive and a negative value of $m$ is impossible
without closing the gap.

We now consider a model with two real fermions propagating in each direction,
so that the mass term has more freedom. This situation occurs, for example, at
the edge of a 2D topological insulator. A gap opens in a magnetic field or in
close contact with a superconductor~\cite{FuKane}. The Hamiltonian is as
follows:
\begin{equation}
\hat{H}=\frac{i}{2}\int\hat{\eta}^T\!(\gamma\partial+M)\hat{\eta}\,dx,\quad
\hat{\eta}=\!\!\left(\begin{array}{@{}c@{}}
\hat{\psi}_{\spinup}+\hat{\psi}_{\spinup}^{\dag}\\
-i(\hat{\psi}_{\spinup}-\hat{\psi}_{\spinup}^{\dag})\\
\hat{\psi}_{\spindown}+\hat{\psi}_{\spindown}^{\dag}\\
-i(\hat{\psi}_{\spindown}-\hat{\psi}_{\spindown}^{\dag})
\end{array}\right)\!
\end{equation}
\begin{equation}
\gamma=\left(\begin{array}{@{}cc@{}}
I & 0\\0 & -I
\end{array}\right),\qquad
M=\left(\begin{array}{@{}cc@{}}
-h_z(i\sigma^y) & m\\-m^{T} & h_z(i\sigma^y)
\end{array}\right),\vspace{4pt}
\end{equation}
\begin{equation}
m=-h_x(i\sigma^y)+h_{y}I-(\Re\Delta)\sigma^x-(\Im\Delta)\sigma^{z}.
\end{equation}
If $h_z=0$, the energy gap is given by the smallest singular value of $m$; it
vanishes at the transition between the ``magnetic'' and ``superconducting''
phase as the function $\det(m)=h_x^2+h_y^2-|\Delta|^2$ passes through
zero. The presence of $h_z$ complicates the matter, but if the spectrum is
gapped, $h_z$ can be continuously tuned to zero without closing the gap. We
will see that, in general, \emph{the mass term can be tuned to anticommute
with $\gamma$}, in which case $M$ consists of two off-diagonal blocks, $m$ and
$-m^T$.

With $n$ modes propagating in each direction, the nondegenerate anticommuting
mass term is given by $m\in\GL(n,\RR)$. This set has two connected components,
hence there are two distinct phases. Note that the set $\GL(n,\RR)$ is
homotopy equivalent to $R_1=O(n)$ (see Table~\ref{tab_Bott}); it provides the
classification of systems with no symmetry for $d=1$ (cf.\
Table~\ref{tab_periodic}). We proceed with a more systematic approach.

\section{Classification principles}

Concrete mathematical problems may be formulated for Dirac operators, band
insulators, or more general systems. Let us set up the framework. We need to
define a set of admissible Hamiltonians and some equivalence relation between
them; the equivalence classes may then be called ``phases''. Continuous
deformation, or \emph{homotopy} is part of the equivalence definition, but it
is not sufficient for a nice classification. A key idea in $K$-theory is that
of \emph{stable equivalence}: when comparing two objects, $X'$ and $X''$, it
is allowed to augment them by some object $Y$. We generally augment by a
trivial system, i.e., a set of local, disjoint modes, like inner atomic
shells. This corresponds to adding an extra flat band on an insulator. It may
be the case that two systems cannot be continuously deformed one to the other,
but such a deformation becomes possible after the augmentation. Thus, the
topological classification of band insulators with an unlimited number of
bands is simpler than in the case of two bands!  Likewise, it is easier to
classify Dirac operators if we do not impose any restriction on the size of
gamma-matrices. The final twist is that $K$-theory deals with
\emph{differences} between objects rather than objects themselves. Thus, we
consider one phase relative to another.

We now give exact definitions for $d=0$ (meaning that the system is viewed as
a single blob). The simplest case is where the particle number is conserved,
but there are no other symmetries. A general free-fermion has this form:
\begin{equation}\label{qhamQ}
\hat{H}=\sum\limits_{j,k}X_{jk} \hat{a}_{j}^{\dag} \hat{a}_{k},
\end{equation}
where $X=(X_{jk})$ is some Hermitian matrix representing electron
hopping. Since we are interested in gapped systems, let us require that the
eigenvalues of $X$ are bounded from both sides, e.g.,
$\Delta\le|\epsilon_{j}|\le E_{\mathrm{max}}$. The following condition is
slightly more convenient:
\begin{equation}\label{evbound}
\alpha\le \epsilon_{j}^2 \le\alpha^{-1},
\end{equation}
where $\alpha\le 1$ is some constant. This class of matrices is denoted by
$C_{0}(\alpha)$, and the corresponding Hamiltonians are called
\emph{admissible}. (Some locality condition will be needed in higher
dimensions, but for $d=0$, this is it.)

The ``spectral flattening'' transformation, $X\mapsto\widetilde{X}=\sgn X$
reduces admissible matrices to a special form, where all positive eigenvalues
are replaced by $+1$, all negative eigenvalues are replaced by $-1$, and the
eigenvectors are preserved. (The matrix element $\widetilde{X}_{jk}$ is,
essentially, the equal-time Green function.) Such special matrices constitute
the set
\begin{equation}
C_{0}(1)=\bigcup_{0\le k\le n}U(n)/(U(k)\times U(n-k)),
\label{C01}
\end{equation}
where $n$ and $k$ are the matrix size and the numbers of $-1$ eigenvalues,
respectively.

We write $X'\approx X''$ (or $X'\Approx{\alpha}X''$ to be precise) if $X'$ and
$X''$ are homotopic, i.e., can be connected by a continuous path within the
matrix set $C_{0}(\alpha)$. It is easy to see that two matrices are homotopic
if and only if they agree in size and have the same number of negative
eigenvalues. For families of matrices, i.e., continuous functions from some
parameter space $\Lambda$ to $C_{0}(\alpha)$, the homotopy classification is
more interesting. For example, consider an integer quantum Hall system on a
torus. The boundary conditions are described by two phases $(\mod 2\pi)$,
therefore the parameter space is also a torus. This family of Hamiltonians is
characterized by a nontrivial invariant, the first Chern
number~\cite{NiuThoulessWu}.

It is clear that $C_{0}(\alpha)$ can be contracted within itself to $C_{0}(1)$
since we can interpolate between the identity map and the spectral flattening:
$X\mapsto f_{t}(X)$, where $t\in[0,1]$,\, $f_{0}(x)=x$,\, $f_{1}(x)=\sgn x$,
and the function $f_{t}$ is applied to the eigenvalues of Hermitian matrix $X$
without changing the eigenvalues. Thus, $C_{0}(\alpha)$ is homotopy equivalent
to $C_{0}(1)$, and we may use the latter set for the purpose of topological
classification.

Let us consider this example (where $X$ is a single matrix or a continuous
function of some parameters):
\begin{equation}
\label{X-X}
Y_0=\left(\begin{array}{@{\,}cc@{\,}}X&0\\0&-X\end{array}\right)
\approx \left(\begin{array}{@{\,}cc@{\,}}0&iI\\-iI&0\end{array}\right)=Y_1,
\end{equation}
The actual homotopy is $Y_t=\cos(t\pi/2)Y_{0}+\sin(t\pi/2)Y_{1}$. Note that
$Y_t^2=1$ since $Y_{0}^2=Y_{1}^2=1$ and $Y_{0}Y_{1}=-Y_{1}Y_{0}$.
Furthermore, $Y_1$ is homotopic to the matrix that consists of $\sigma^z$
blocks on the diagonal; such matrices will be regarded as \emph{trivial}. This
example shows that any admissible system ($X$) is effectively canceled by its
particle-hole conjugate ($-X$), resulting in a trivial system. That is always
true for free-fermion Hamiltonians, with any symmetry, in any dimension.

\emph{Equivalence} between admissible matrices is defined as follows:
\begin{equation}
X'\sim X''\quad \textrm{if}\quad X'\oplus Y\approx X''\oplus Y\,\ \textrm{for
some}\,\ Y,
\end{equation}
where $\oplus$ means building a larger matrix from two diagonal
blocks. Without loss of generality, we may assume that $Y$ is trivial. Indeed,
if $X'\oplus Y\approx X''\oplus Y$, then $X'\oplus Y\oplus(-Y)\approx
X''\oplus Y\oplus(-Y)$, and we have seen that $Y\oplus(-Y)$ is homotopic to a
trivial matrix.

The \emph{difference class} $d(A,B)$ of two same-sized matrices is represented
by the pair $(A,B)$ up to this equivalence relation:
\begin{equation}
(A',B')\sim (A'',B'')\quad \textrm{if}\quad
A'\oplus B''\sim A''\oplus B'.
\end{equation}
Note that the the matrix sizes in different pairs need not be the same. Since
$(A,B)\sim(A\oplus(-B),\,B\oplus(-B))$, it is sufficient to consider pairs
where the second matrix is trivial. Thus, the equivalence class of $(A,B)$ is
given by a single integer, $k=k(A)-k(B)$, where $k(\cdots)$ denotes the number
of negative eigenvalues. Since $B$ is trivial, $k(B)$ equals half the matrix
size, $n=2s$. Hence, $k(A)=s+k$.

To characterize the difference between two \emph{families} of matrices
parametrized by $\Lambda$, one needs to consider functions from $\Lambda$ to
the classifying space $C_{0}$:~\footnote{Here $\lim_{m\to\infty}$ is a
so-called \emph{direct limit}: the unitary cosets for smaller $m$ are mapped
into ones for larger $m$.}
\[
C_{0}=\bigcup_{k\in\ZZ}\lim_{s\to\infty}U(2s)/(U(s+k)\times U(s-k)).
\]
It is the same space as in Table~\ref{tab_Bott}. The Abelian group of
difference classes ($=$ homotopy classes of functions $\Lambda\to C_{0}$) is
denoted by $K_{\CC}^{0}(\Lambda)=\pi(\Lambda,C_{0})$.

\section{Symmetries and Clifford algebras}

In this section, we complete the $d=0$ classification. Since the particle
number is not generally conserved, we will use the Hamiltonian $H_{A}$ given
by a real skew-symmetric matrix $A$ (see Eq.~(\ref{quadH})). To generalize
some arguments of the previous section, let us also define the trivial
Hamiltonian:
$\hat{H}_{\mathrm{triv}}
=\sum_{j}\bigl(\hat{a}_{j}^{\dag}\hat{a}_{j}-\frac{1}{2}\bigr)
=\hat{H}_{Q}$, where
\begin{equation}\label{Q}
Q=\left(\begin{array}{@{}ccccc@{}}
  0 & 1 &&&\\ -1 & 0 &&&\\
  && 0 & 1 &\\ && -1 & 0 &\\ &&&&\ddots \end{array}\right)\,.
\end{equation}

The eigenvalues of $A$ come in pairs $(+i\epsilon_{j},-i\epsilon_{j})$, where
$\epsilon_{j}$ are positive and satisfy inequality~(\ref{evbound}). Replacing
$A$ with $\widetilde{A}=-i\sgn(iA)$ takes $\epsilon_{j}$ to $1$. The matrix
$\widetilde{A}$ can be represented as $SQS^{-1}$, where $S\in O(2n)$. However,
this representation is not unique since $S$ can be multiplied on the right by
any orthogonal matrix that commutes with $Q$. Such matrices form a subgroup of
$O(2n)$ that may be identified with $U(n)$. Thus, the set of matrices
$\widetilde{A}$ (i.e., real skew-symmetric matrices with $\pm i$ eigenvalues)
is equal to $O(2n)/U(n)$. Let us take the $n\to\infty$ limit by identifying
$\widetilde{A}$ with $\widetilde{A}\oplus Q$ (where the size of $Q$ can be any
even number). The result is listed in Table~\ref{tab_Bott} as the classifying
space $R_2$:
\[
R_2=\lim_{n\to\infty} O(2n)/U(n).
\]
The set $R_2$ has two connected components, which are distinguished by the
value of $\sgn(\Pf A)=\Pf\widetilde{A}=\det S=\pm1$. The physical meaning of
this invariant is the \emph{fermionic parity} $(-1)^{\hat{N}}$ in the ground
state, where $\hat{N}=\sum_{j}\hat{a}_j^{\dag}\hat{a}_j$ is the particle
number. Note that $\hat{N}$ is conserved $(\mod 2)$.

The condition that $\hat{N}$ is conserved as an integer is equivalent to a
$U(1)$ symmetry. In this case, the creation-annihilation expression of
Hamiltonian~(\ref{quadH}) should not contain terms like $\hat{a}_j\hat{a}_k$
or $\hat{a}_k^{\dag}\hat{a}_j^{\dag}$. This is a good point to note that the
approach based on free-fermion Hamiltonians is fundamentally incomplete since
it cannot distinguish between the full $U(1)$ group and its $\ZZ_{4}$
subgroup, which is generated by the transformation $\hat{a}_{j}\mapsto
i\hat{a}_{j}$. Let us assume for a moment that the actual symmetry is
$\ZZ_4$. Then terms like $\hat{a}_1\hat{a}_2\hat{a}_3\hat{a}_4$ are allowed in
principle, but not in a free-fermion Hamiltonian. Therefore topological
invariants of noninteracting systems may not be preserved in the presence of
interactions. In the following example, the number of particle-occupied states
changes by $4$ by a continuous path through an interacting phase:
\[
\hat{H}(t)=\cos(\pi t)\sum_{j=1}^{4}\hat{a}_{j}^{\dag}\hat{a}_{j}+\sin(\pi t)
\bigl(\hat{a}_1\hat{a}_2\hat{a}_3\hat{a}_4+\textrm{h.c.}).
\]
Note that the ground state remains non-degenerate for all values of $t$. On
the other hand, a homotopy like that is only possible if the interaction term
exceeds the energy gap at some point. Thus, the noninteracting topological
classification is generally stable to weak interactions, but not to strong
ones. In the $U(1)$ case, it is absolutely stable though (at least, for
$d=0$). We now set this discussion aside and proceed with the noninteracting
case.

It is easy to see that the Hamiltonian~(\ref{quadH}) is $U(1)$ invariant if
and only if the matrix $A$ commutes with $Q$ (see Eq.~(\ref{Q})). Another
possible symmetry is time-reversal invariance. It can be expressed by an
antiunitary operator $\hat{T}$ acting in the Fock space; this action is
defined as follows:
\[
\hat{T}i\hat{T}^{-1}=-i,\quad
\begin{array}{r@{}l@{\quad}r@{}l}
\hat{T}\hat{a}_{j\spinup}\hat{T}^{-1}&=\hat{a}_{j\spindown},
&\hat{T}\hat{a}_{j\spinup}^{\dag}\hat{T}^{-1}&=\hat{a}_{j\spindown}^{\dag},
\\[3pt]
\hat{T}\hat{a}_{j\spindown}\hat{T}^{-1}&=-\hat{a}_{j\spinup},
&\hat{T}\hat{a}_{j\spindown}^{\dag}\hat{T}^{-1}&=-\hat{a}_{j\spinup}^{\dag}.
\end{array}
\]
Converting $\hat{a}_{1\spinup}, \hat{a}_{1\spinup}^{\dag},
\hat{a}_{1\spindown}, \hat{a}_{1\spindown}^{\dag},\ldots$ into
$\hat{c}_1,\hat{c}_2,\hat{c}_3,\hat{c}_4,\ldots$, we obtain a relation of the
form $\hat{T}\hat{c}_{m}\hat{T}^{-1}=\sum_{l}T_{lm}\hat{c}_{l}$, where the
matrix $T$ consists of $4\times4$ blocks:
\begin{equation}
T=\left(\begin{array}{@{}ccccc@{}}
  0 & 0 &-1 & 0 &\\ 0 & 0 & 0 & 1 &\\
  1 & 0 & 0 & 0 &\\ 0 &-1 & 0 & 0 &\\ &&&&\ddots \end{array}\right)\,.
\end{equation}
The $\hat{T}$-invariance of the Hamiltonian is equivalent to the condition
$TA=-AT$.

Let us describe a common algebraic structure that is applicable to three
symmetry types: no symmetry, $T$ only, and $T$ and $Q$. First, note these
identities:
\begin{equation}
T^2=Q^2=-1, \qquad TQ=-QT.
\end{equation}
It is convenient to introduce some new notation: $e_1=T$,\, $e_2=QT$. Note
that $e_2$ anticommutes with $A$ if both the $T$ and $Q$ symmetries are
present. Let us also use $\widetilde{A}=-i\sgn(iA)$ instead of $A$. Then we
have the following characterization:
\begin{description}
\item[No symmetry:]\quad $\widetilde{A}^2=-1$;
\item[$T$ only:]\quad $e_1^2=\widetilde{A}^2=-1$,\quad
$e_1\widetilde{A}=-\widetilde{A}e_1$;
\item[$T$ and $Q$:]\quad \begin{tabular}[t]{@{}l@{}}
$e_1^2=e_2^2=\widetilde{A}^2=-1$,\\[3pt]
$e_1e_2=-e_2e_1$,\quad $e_j\widetilde{A}=-\widetilde{A}e_j$\quad ($j=1,2$).
\end{tabular}
\end{description}
The pattern is pretty obvious. We have $p$ predefined matrices
$e_{1},\dots,e_{p}$\, ($p=0,1,2$) satisfying Clifford algebra
relations (see exact definition below) and look for all possible choices of
another Clifford generator $e_{p+1}=\widetilde{A}$.

The \emph{(real) Clifford algebra} $\Cliff^{p,q}$ is generated by elements
$e_1,\dots,e_{p+q}$ satisfying these relations:\footnote{An alternative
notation is also used, where the positive generators ($e_j^2=1$) are listed
first and the parameters $p$ and $q$ are swapped.}
\begin{equation}
\begin{array}{c}
e_1^2=\dots=e_p^2=-1,\quad\; e_{p+1}^2=\dots=e_{p+q}^2=1,\\[3pt]
e_je_k=-e_ke_j\quad (j\not=k).
\end{array}
\end{equation}
All Clifford algebras can be described in terms of the 3 simple algebras with
real coefficients: $\RR$ (real numbers), $\CC$ (complex numbers), and $\HH$
(quaternions).  For example, $\Cliff^{0,1}$ is isomorphic to $\RR\oplus\RR$
since it consists of linear combinations of two complementary projectors,
$\frac{1}{2}(1\pm e_1)$. The algebra $\Cliff^{1,0}$ can be identified with
$\CC$ by mapping the negative generator $e_{1}$ to $i$. Furthermore,
$\Cliff^{0,2}\cong\Cliff^{1,1}\cong\RR(2)$ (the algebra of real $2\times2$
matrices where the Clifford generators are mapped to $\sigma^z$, $\sigma^x$ or
to $\sigma^z$, $i\sigma^y$, respectively), and $\Cliff^{2,0}\cong\HH$. For
more details on Clifford algebras and their use in $K$-theory, see
Refs.~\cite{AtiyahBottShapiro,Karoubi}.

In the problem at hand, the Clifford generators act in the mode
space.\footnote{In comparison, the Majorana operators $\hat{c}_l$ generate a
(complex) Clifford algebra acting in the Fock space.} Thus, we deal with
Clifford algebra \emph{representations} such that $e_1,\dots,e_p$ and
$e_{p+1},\dots,e_{p+q}$ are represented by real skew-symmetric and real
symmetric matrices, respectively. To classify free-fermion Hamiltonians,
we consider representations of $\Cliff^{p+1,0}$ with fixed action of
$e_1,\dots,e_p$; we call that the ``Clifford extension problem with $p$
negative generators''.

For technical reasons, it is convenient to reformulate the problem in terms of
positive generators. To this end, we will employ the isomorphism
$\Cliff^{0,p+2}\cong\Cliff^{p,0}\otimes\RR(2)$, which may be defined as
follows:
\begin{equation}
\begin{array}{c}
e_{j}\mapsto e_{j}'\otimes(i\sigma^{y})\quad \textrm{for}\ j=1,\dots,p,\\[3pt]
e_{p+1}\mapsto I\otimes\sigma^{z},\quad e_{p+2}\mapsto I\otimes\sigma^{x}.
\end{array}
\end{equation}
Representations of the algebra $\calA\otimes\RR(n)$ (for any $\calA$) have
very simple structure, namely, $E\otimes\RR^n$, where the first factor is some
representation of $\calA$ and the second comes with the standard action of the
matrix algebra $\RR(n)$. Thus, $\calA$ and $\calA\otimes\RR(n)$ have the same
representation theory (i.e., their representations are in a natural one-to-one
correspondence); such algebras are called \emph{Morita equivalent}. Up to
Morita equivalence, $\Cliff^{p,q}$ only depends on $p-q\bmod 8$.

We conclude that the classification of free-fermion Hamiltonians with $p$
negative Clifford symmetries is equivalent to the extension problem with
$q=p+2$ positive generators. That is to say, we need to find all possible
actions of $e_{q+1}$\, ($e_{q+1}^2=1$) if the action of $e_1,\dots,e_q$ is
fixed.  In $K$-theory, the problem is formulated in terms of \emph{difference
objects} $(E,F,w)$, where $E$, $F$ are representations of $\Cliff^{0,q+1}$ and
$w$ is a linear orthogonal map that identifies them as $\Cliff^{0,q}$
representations, see~\cite{Karoubi}. Without loss of generality, we may fix
$F$ to be a sum of several copies of the regular representation (which
corresponds to a trivial Hamiltonian) and $w$ the identity map. Such
difference objects form the classifying space $R_q$ (see
Table~\ref{tab_Bott}). The Abelian group of equivalence classes of difference
objects parametrized by $\Lambda$ is
$K_{\RR}^{0,q}(\Lambda)=\pi(\Lambda,R_q)$. It is isomorphic to the
conventional real $K$-group $K_{\RR}^{-q}(\Lambda)$, which is also denoted by
$KO^{-q}(\Lambda)$. In the special case where $\Lambda=\pt$ (a single point),
we get $K_{\RR}^{-q}(\pt)=\pi_0(R_q)$.

\section{Classification for arbitrary $d$}

We begin with a short summary, focusing on the symmetry classes that
correspond to real $K$-theory. It is natural to distinguish three cases:
\begin{enumerate}
\item \emph{Continuous} free-fermion Hamiltonians are classified by
$\widetilde{K}_{\RR}^{-q}(\Bar{S}^d)=\pi_0(R_{q-d})$, where $\Bar{S}^d$
represents the momentum space (see below). Sufficient insight can be gained by
considering Dirac operators. This setting is actually more general than one
might expect: gapped Hamiltonians in the momentum space are topologically
equivalent to nondegenerate mass terms that anticommute with a fixed Dirac
operator. Long-range disorder may be described by \emph{textures} of the mass
term varying in space, i.e., continuous functions $M:\RR^{d}\to R_{q-d}$.
\item \emph{Band insulators} are characterized by the momentum space
$\Bar{\TT}^{d}$, hence the classification is given by
$K_{\RR}^{-q}(\Bar{\TT}^d)$. This Abelian group includes
$\pi_0(R_{q-d})$ as a direct summand, but there is some extra piece (cf.\
``weak topological insulators'').
\item \emph{Arbitrary local discrete systems} under the energy gap or
localization condition. (``Local'' means that the electron hopping is
short-ranged. The gap condition is stronger than the localization, but the
problem for the localized case can be reduced to that for the gapped case.)
Realizations of short-range disorder fall into this category. The
classification of such general systems is exactly the same as for Dirac
operators, due to the following\vspace{2pt}

\textbf{Theorem:} \emph{Any gapped local free-fermion Hamiltonian in $\RR^d$
is equivalent to a texture.}\vspace{2pt}

(That is the key technical result, but I cannot explain it in any detail in
such a short note.) Discrete systems on a compact metric space $L$ are
classified by the $K$-homology group $K^{\RR}_{q}(L)$.
\end{enumerate}

\subsection{Continuous systems and Dirac operators}

The Hamiltonian of a translationally invariant systems can be written in the
momentum representation:
\begin{equation}
\hat{H}=\frac{i}{4}\sum_{\V{p}}\sum_{j,k}A_{jk}(\V{p})
\hat{c}_{-\V{p},j}\hat{c}_{\V{p},k},
\end{equation}
where $j$ and $k$ refer to particle flavors. The matrix $A(\V{p})$ is
skew-Hermitian but not real; it rather satisfies the condition
$A_{jk}(\V{p})^{*}=A_{jk}(-\V{p})$. By abuse of terminology, such matrix-valued
functions are called ``functions from $\Bar{\RR}^d$ to real skew-symmetric
matrices'', where $\Bar{\RR}^{d}$ is the usual Euclidean space with the
involution $\V{p}\leftrightarrow-\V{p}$ (cf.~\cite{Kreality}). The symmetry is
defined by some Clifford generators represented by real matrices whose
action does not depend on $\V{p}$. As described in the previous section, we
can turn negative generators to positive and replace $A(\V{p})$ by another
Clifford generator $e_{q+1}(\V{p})$. While the matrices $e_{1},\dots,e_{q}$
are real symmetric, $e_{q+1}$ is Hermitian and satisfies the condition
$e_{q+1}(\V{p})^{*}=e_{q+1}(-\V{p})$.

A reasonable classification can be developed when the asymptotics of
$A(\V{p})$ for $|\V{p}|\to\infty$ is fixed. We may identify the infinity in
the momentum space with the boundary of a large ball,
$\partial\Bar{B}^{d}$. Thus, the difference between two phases may be
characterized by an element of the relative $K$-group
\begin{equation}\label{contK}
K_{\RR}^{0,q}(\Bar{B}^{d},\partial\Bar{B}^{d})
=\widetilde{K}_{\RR}^{0,q}(\Bar{S}^{d}) \cong \pi_{0}(R_{q-d}).
\end{equation}
Here we have used the isomorphism~\cite{Karoubi}
\begin{equation}
\widetilde{K}_{\RR}^{p,q}(X)\cong\widetilde{K}_{\RR}^{0}(S^{r}X)\qquad
(r=q-p\bmod 8),
\end{equation}
and the $(1,1)$ periodicity~\cite{Kreality}:
\begin{equation}
\widetilde{K}_{\RR}^{0}(S\Bar{S}X)\cong\widetilde{K}_{\RR}^{0}(X),
\end{equation}
where $S$ denotes the suspension.

The group $\pi_{0}(R_{q-d})\cong K_{\RR}^{d,q}(\pt)$ on the right-hand side of
Eq.~(\ref{contK}) has a concrete physical interpretation. It classifies the
nondegenerate mass terms $M$ in the real self-adjoint Dirac operator
$D=\sum_a\gamma_{a}\partial_{a}+M$, where $\gamma_{a}$ are skew-symmetric, $M$
is symmetric, and
\begin{equation}
\gamma_{a}\gamma_{b}+\gamma_{b}\gamma_{a}=-\delta_{ab},\qquad
\gamma_{a}M=-M\gamma_{a}.
\end{equation}
(Replacing $M$ with $\widetilde{M}=\sgn M$, we can achieve that $M^{2}=1$.) In
addition, we assume that $\gamma_{1},\dots,\gamma_{d}$ and $M$ anticommute
with the symmetry generators $e_{1},\dots,e_{q}$. Thus, the gamma-matrices
play the role of Clifford symmetries with opposite sign; they effectively
cancel the actual symmetries. Note that those new ``symmetries'' do not entail
any conservation laws. Our argument only implies that any continuous spectrum
is \emph{equivalent} (up to an augmentation and homotopy) to a Dirac spectrum
that has the additional symmetries.

\subsection{Discrete systems}

Let us consider the Hamiltonian~(\ref{quadH}), where each mode $j$ is
associated with a site, or point $\Vr{j}$ in the real space. There may be
several modes per site; symmetries (if any) act independently on each site. We
assume that the Hamiltonian is \emph{$r$-local} (i.e., $A_{jk}=0$ if the
distance between $\Vr{j}$ and $\Vr{k}$ is greater than $r$) and that it is
\emph{$\alpha$-gapped} (i.e., the eigenvalues $\epsilon_j$ of $iA$ satisfy
inequality~(\ref{evbound})). Under these conditions, the matrix element
$\widetilde{A}_{jk}$ decays very fast as the distance between $j$ and $k$ goes
to infinity, which is a sign of localization. Conversely, if we start with the
matrix $\widetilde{A}$ (such that $\widetilde{A}^2=-1$ and
$\widetilde{A}_{jk}$ decays fast enough) and set all the elements for
$|\Vr{j}-\Vr{k}|>r'$ to zero, we will obtain a gapped local matrix. Both
transformations can be done continuously, which roughly shows that the set of
localizing Hamiltonians is contractible within itself to the set of gapped
Hamiltonians (up to a change of controlling parameters).  Thus, we may stick
with the gapped case without any loss of generality.

Using the standard trick, we replace $A$ with a real \emph{symmetric} matrix
$X$ that is $r$-local, $\alpha$-gapped, and anticommutes with $q$
\emph{positive} Clifford symmetries. The above-mentioned theorem pertains to
such matrices. Here, we only discuss it at the physical level. The texture
corresponding to the matrix $X$ is constructed algorithmically, albeit in a
contrived fashion. The procedure is local, with a characteristic radius
$l=cr$, where $c$ depends on $d$ and $\alpha$. The number of Dirac modes
needed is $l^d$ (for localized systems, it's the localization volume).  To
calculate $M(\V{r})$, we only look at the $l$-neighborhood of point $\V{r}$,
and $M$ doesn't vary much at distances smaller than $l$. We may then
discretize the Dirac operator on a fine grid, with $\gamma_a$ and $M$ rescaled
properly so as to keep the $\alpha$ parameter fixed. Thus, we obtain an
$r'$-local, $\alpha$-gapped matrix $X'$, where $r'$ is arbitrary small. The
equivalence between $X$ and $X'$ involves an augmentation and a homotopy,
where $r$ may increase by a constant factor before it shrinks down to $r'$.

This theorem implies that the boundary between two phases must carry some
gapless modes. Indeed, each phase may be characterized by the mass term
$M(\V{r})$ computed at any point away from the boundary. Since the phases are
different, the two mass terms, $M(\V{r}_1)$, $M(\V{r}_2)$ belong to different
connected components of the classifying space. But if the boundary between the
phases were gapped, we could make the whole system into a continuous texture,
and thus $M(\V{r}_1)$ and $M(\V{r}_2)$ would belong to the same component ---
a contradiction.

A gapped local system on a compact metric space $L$ (say, a manifold with or
without boundary) is characterized by a $K$-homology class $\xi\in
K^{\RR}_{q}(L)$, where $q$ is defined $(\mod 8)$. $K$-homology (see
e.g.~\cite{HigsonRoe}) and the related noncommutative geometry~\cite{Connes}
are advanced subjects, but the basic intuition is rather simple. Let us
consider systems with no symmetry ($q=2$) on the two-dimensional torus,
$L=\TT^2$. Such systems are trivially characterized by the number of fermions
in the ground state, $\xi_{0}\in\ZZ_{2}$. Now imagine a closed Majorana chain
winding around the torus. It defines a homology class $\xi_1\in
H_1(\TT^2;\ZZ_2)$, which is a topological invariant for gapped local
systems. It can be measured by cutting the torus along some cycle $c$ and
counting edge modes $(\mod 2)$. Or one can flip the sign of all matrix
elements $A_{jk}$ spanning across the cut and see how $\xi_0$ changes:
$\xi_1(c)=\xi_0(+)\,\xi_{0}(-)$. If the torus is filled with a $p_x+ip_y$
superconductor, the system has a nontrivial two-dimensional invariant,
$\xi_2\in\ZZ$. But if $\xi_2$ is odd, then the properties of the 1D invariant
change: $\xi_1$ is not a homology class, but rather, a spin structure. Indeed,
\begin{equation}
\xi_0(++)\,\xi_0(+-)\,\xi_0(-+)\,\xi_0(--)=(-1)^{\xi_2},
\end{equation}
where $\pm$ refers to the sign of matrix elements across two basis cycles. In
general, the definition of low-dimensional invariants (except in dimension
$0$) depends on the higher-dimensional ones.\footnote{The term ``invariant''
is used in a sloppy way, but one can rigorously define the range of $\xi_s$,
assuming that $\xi_{s+1}=\dots=\xi_{d}=0$. It is the Abelian group
$E_{s,q-s}^{\infty}$ of the $K$-homology spectral sequence.} The $K$-homology
class includes all.

\subsection{Band insulators}

The main difference from continuous systems is that the momentum space is
$\Bar{T}^{d}$. Since there is no need to fix the spectrum at infinity, the
classification is given by the absolute $K$-group
$K_{\RR}^{0,q}(\Bar{T}^{d})\cong K_{\RR}^{-q}(\Bar{T}^{d})$. The band
structure analysis in Refs.~\cite{KaneMele1,MooreBalents,FuKaneMele} and
others offers a concrete view of that group in certain cases. Unfortunately,
the momentum space picture is non very intuitive. To understand and calculate
the group $K_{\RR}^{-q}(\Bar{T}^{d})$, we relate it to $K$-homology of the
real-space torus by means of the Baum-Connes isomorphism for $\ZZ^{d}$ (a
$K$-theory analogue of the Fourier transform). Then we apply the Poincare
duality. Thus,
\begin{equation}
\begin{array}{r@{}l}
K_{\RR}^{-q}(\Bar{\TT}^d)
&\cong K^{\RR}_{q}(\TT^d)\cong K_{\RR}^{d-q}(\TT^d)
\\[4pt]
&\cong \pi_{0}(R_{q-d})\oplus \widetilde{K}_{\RR}^{d-q}(\TT^d).
\end{array}
\end{equation}
The first term is the same as before, but the last one is new. It further
splits, though not canonically:
\begin{equation}
\widetilde{K}_{\RR}^{d-q}(\TT^d)\cong
\bigoplus_{s=0}^{d-1}{d\choose s}\,\pi_{0}(R_{q-s}).
\end{equation}
For 3D $T$-invariant insulators, i.e., $d=3$,\, $q=4$, we get:
\begin{equation}
\widetilde{K}_{\RR}^{-1}(\TT^3)
\cong \ZZ\oplus 3\ZZ_2.
\end{equation}
The $\ZZ$ term is the number of (Kramers degenerate) valence bands,
whereas $3\ZZ_2$ pertains to ``weak topological insulators''.

\section{The effect of interaction}

Topological properties of gapped local free-fermion systems are mostly
understood. The big open question is how the classification is changed by
interactions, e.g., whether different free-fermion phases can be deformed one
to another through an interacting phase without closing the gap. In some
cases, e.g., the integer quantum Hall effect and chiral 2D superconductors,
the topological invariants are related to physical properties that are
well-defined in the presence of interactions (namely, the Hall conductivity
and the chiral central charge, which determines the edge energy
current~\cite{KaneFisher,hexagon}). The Kramers degeneracy analysis of
vortex-bound states demonstrates the stability of 2D topological
insulators~\cite{KaneMele1,Z2pump} and
$(p_x\!+\!ip_y)\spinup+(p_x\!-\!ip_y)\spindown$
superconductors~\cite{classSuper}.

However, the free-fermion classification is unstable for 1D systems with the
unusual $T$ symmetry: $\hat{T}^2=1$ instead of
$\hat{T}^{2}=(-1)^{\hat{N}}$. For a concrete model, consider the Majorana
chain and its variations, where $\hat{T}$ acts on odd sites by
$\hat{T}\hat{c}_{j}\hat{T}^{-1}=-\hat{c}_{j}$ so that terms like
$i\hat{c}_{j}\hat{c}_{k}$ are only allowed between sites of different
parity. In the free-fermion setting, this symmetry is described by one
positive Clifford generator, hence $p=-1$,\,\, $q=p+2=1$, and for $d=1$ we get
a topological invariant $k\in\pi_{0}(R_{q-d})=\ZZ$. For example, the usual
phase transition in $8$ parallel Majorana chains is characterized by
$k=8$. But in this particular case, the two phases are actually connected
through an interacting phase~\cite{Majorana8}.
\medskip

\section{Acknowledgments}
I am grateful to Andreas Ludwig and Shinsey Ryu for teaching me about
$^{3}$He-$B$ and $(p_x\!+\!ip_y)\spinup+(p_x\!-\!ip_y)\spindown$ and helping
to fit these phases into the periodic table. I also thank John Preskill,
Michael Freedman, John Roe, Charles Kane, and Grigori Volovik for stimulating
discussions. This research is supported in part by NSF under grant No.\
PHY-0456720.

\end{document}